\documentclass[sigconf]{acmart}
\pdfoutput=1

\usepackage{url}

\begin{document}
\title{TYDR -- Track Your Daily Routine.\\Android App for Tracking Smartphone Sensor and Usage Data}

\author{Felix Beierle, Vinh Thuy Tran}
\affiliation{%
	\institution{Service-centric Networking\\Technische Universit\"at Berlin\\Telekom Innovation Laboratories}
	\city{Berlin}
	\country{Germany}
}
\email{beierle@tu-berlin.de,vinh.t.tran@campus.tu-berlin.de}

\author{Mathias Allemand}
\affiliation{%
	\institution{Department of Psychology\\University of Zurich}
	\city{Zurich}
	\country{Switzerland}
}
\email{m.allemand@psychologie.uzh.ch}

\author{Patrick Neff, Winfried Schlee}
\affiliation{%
	\institution{Clinic and Policlinic for Psychiatry and Psychotherapy\\University of Regensburg}
	\city{Regensburg}
	\country{Germany}
}
\email{patrick.neff@uzh.ch,winfried.schlee@tinnitusresearch.org}

\author{Thomas Probst}
\affiliation{%
	\institution{Department for Psychotherapy and Biopsychosocial Health\\Danube University Krems}
	\city{Krems}
	\country{Austria}
}
\email{thomas.probst@donau-uni.ac.at}

\author{R\"udiger Pryss}
\affiliation{%
	\institution{Institute of Databases and Information Systems\\Ulm University}
	\city{Ulm}
	\country{Germany}
}
\email{ruediger.pryss@uni-ulm.de}

\author{Johannes Zimmermann}
\affiliation{%
	\institution{Psychologische Hochschule Berlin}
	\city{Berlin}
	\country{Germany}
}
\email{j.zimmermann@psychologische-hochschule.de}

\renewcommand{\shortauthors}{F. Beierle et al.}

\begin{abstract}

We present the Android app TYDR (Track Your Daily Routine)
which tracks smartphone sensor and usage data and
utilizes standardized psychometric personality questionnaires.
With the app, we aim at collecting data for researching
correlations between the tracked smartphone data and the user's personality
in order to predict personality from smartphone data.
In this paper,
we highlight our approaches in addressing the challenges in developing
such an app.
We optimize the tracking of sensor data by assessing the
trade-off of size of data and battery consumption
and granularity of the stored information.
Our user interface is designed to incentivize
users to install the app and fill out questionnaires.
TYDR processes and visualizes the tracked sensor and usage data
as well as the results of the personality questionnaires.
When developing an app that will be used in psychological studies,
requirements posed by ethics commissions / institutional review boards and data protection officials
have to be met.
We detail our approaches concerning those requirements regarding the
anonymized storing of user data,
informing the users about the data collection,
and enabling an opt-out option.
We present our process for anonymized data storing
while still being able to identify individual users
who successfully completed a psychological study with the app.

\end{abstract}



\copyrightyear{2018} 
\acmYear{2018} 
\setcopyright{rightsretained} 
\acmConference[MOBILESoft '18]{MOBILESoft '18: 5th IEEE/ACM International Conference on Mobile Software Engineering and Systems}{May 27--28, 2018}{Gothenburg, Sweden}
\acmBooktitle{MOBILESoft '18: 5th IEEE/ACM International Conference on Mobile Software Engineering and Systems, May 27--28, 2018, Gothenburg, Sweden}\acmDOI{10.1145/3197231.3197235}
\acmISBN{978-1-4503-5712-8/18/05}


\maketitle

\section{Introduction}
\label{sec:intro}

Context-aware applications consider the context the user
is currently in.
This usually includes factors like the location, weather,
or time.
As previous research suggests, higher level information like the user's
personality could be beneficial to improve
context-awareness in mobile applications.
This has been shown for 
areas like mobile health \cite{HalkoPersonalityPersuasiveTechnology2010a} or
mobile social networking \cite{beierle_towards_2017}.
Assessing the user's personality is a tedious task that is usually done
by filling out questionnaires.
As researchers in computer science and psychology,
we aim at predicting the personality of a user by their
smartphone usage behavior and sensor data.
For the purpose
of collecting data for training a prediction model,
we developed the app TYDR (Track Your Daily Routine).
It tracks smartphone sensor and usage data
and queries the user with standardized psychometric questionnaires
that yield measurements of personality traits.

There are several things to consider when developing such an app.
The more users we can attract, the more reliable our results will be.
The app should have an appealing interface and should offer some feature
for the user.
As we are developing for a mobile device, we have to consider the
restrictions these devices pose, like battery and space limitations.
The concrete study we plan includes a questionnaire that has to be filled out daily.
Finding users for this might have to be incentivized externally.
When conducting psychological studies, it is common that users get paid or are compensated
with university credit points.
For such compensation, it is important to know which users successfully completed the study.
At the same time however, linking the sensitive data that TYDR tracks
to identifying information about the user is highly undesirable because of privacy concerns.
To summarize, the requirements for an app to conduct our study, are:
(1) be appealing to attract users,
(2) consider restrictions posed by developing for mobile devices, and
(3) privacy, especially the collection of anonymized data while still being able to tell which
individual users completed the study successfully.

In this paper, we present TYDR
and highlight the challenges in developing an Android app for the defined use case.
In the following, we
present our design and optimizations for tracking sensor data (Section \ref{sec:tracking}).
In Section \ref{sec:userinterface}, we detail
how TYDR makes the data tracking its core feature
by processing and visualizing it for the user.
In Section \ref{sec:privacy}, we describe
the measures we implemented for privacy protection and data anonymization.

\section{Sensor Data Tracking}
\label{sec:tracking}

Google released the Google Awareness API\footnote{\url{https://developers.google.com/awareness/}}
that offers developers to retrieve different context data through one API
(time, location, places, beacons, headphones, activity, weather).
There are two ways of retrieving data:
\emph{Fence API} and \emph{Snapshot API}.
A snapshot yields current data from the seven sources.
Through the Fence API, the developer can register listeners and receives
a callback when the desired conditions are met.
These two approaches are useful for the development of context-aware applications
that either instantly need the current context of the user
or want to be notified when the user is in a specific context.
Aiming at tracking the user's context, we have to go beyond
what the Google Awareness API offers.
In TYDR, we track: location, weather, ambient light sensor, accelerometer, activity, steps,
phone un-/lock, headphone un-/plug, battery and charging, Wifi, Bluetooth, calls metadata,
music metadata, photos metadata, notifications metadata, app usage, and app traffic.

For some of the listed data sources we want to track,
a passive, listener-based approach is possible:
the music that is played is broadcast by (most) music player apps.
We can just register a listener and track the played back music
whenever the user is listening to it \cite{beierle_privacy-aware_2016}.
The same approach can be used for tracking the usage of the phone:
registering listeners for locking and unlocking events
enables us to track when the user was (most likely) interacting with the phone.

Besides such a listener-based approach, in some cases we have to do
periodical tracking.
Given the related permission, the Android system offers
information about which apps were used for what number of seconds.
Android also offers information about the traffic each app caused.
Such data about the user is especially meaningful when we have it on
a fine-granular level.
For example:
having one data point per week that indicates for how long a user used a specific
app would probably yield less interesting insights than knowing
the app usage duration for each hour.
As there are no listeners available for querying such app statistics, it is necessary
to schedule periodic queries.

For periodically tracking context data, there is the trade-off between the
\begin{enumerate}
	\item amount of data we have to store,
	\item battery-consumption, and
	\item level of detail of the information.
\end{enumerate}
In the most extreme case, the frequency with which we query context
data is very high, in order to increase the level of detail
for quickly changing data.
Then we -- and the user -- have to accept a higher battery consumption.
An example for this quickly changing data is that of the light sensor
or the accelerometer.
We did some further optimizations, for example, by
only tracking accelerometer data when we detect movement
indicated by the step count we track.
Although the light sensor only yields data when the phone is unlocked,
the frequency with which it yields data is still so high that
it would use too much space.
Furthermore, such detailed light sensor data would not be
that useful for our research purpose.
What could be interesting is to check whether the user is in
a dark or in a well-lit place.
To reduce the amount of data we store from the light sensor,
we divided the possible range of light sensor values into several segments
and only store changes between the segments.
To counteract the effects of rapid changes between the
limits of the defined segments, we implemented a hysteresis.
The introduced inaccuracy is negligible for our research purpose.

\section{User Interface}
\label{sec:userinterface}
In this section, we describe TYDR's user interface.
In Section \ref{sec:tiles}, we show the main screen where
the users can see the data that is tracked about them.
Section \ref{sec:notification} is about the permanent notification
and how it can be configured.
Section \ref{sec:questionnaire} introduces the mobile questionnaire
interface that will be used in psychological studies.

\subsection{Main Screen}
\label{sec:tiles}
Figure \ref{fig:main-menu} shows the main screen.
The tile-based design gives the user an immediate overview
of the data for the current day.
Each tile can be touched to slide open a bigger tile with
a weekly view of the data.
In the figure, the user views the weekly data for her/his phone
usage times.
The red tile with the "Grant Permission" button
shows the number of locations the user stayed at.
As the location permission was not granted, the tile cannot display
data and displays the button instead.
Not shown in the screenshot due to space limitations are two additional tiles
related to the number of notifications per app
and the number of photos taken.

\subsection{Customizable Permanent Notification}
\label{sec:notification}

In order to improve the chances of TYDR not being stopped by 
task cleaner apps, we implemented a permanent notification.
It runs
those services in the foreground that track data with a high frequency.
To make it appealing, we followed the same approach as for the main screen:
show the user meaningful and informative figures
based on processed tracking data.

\begin{figure}[p]
	\centering
	\includegraphics[width=0.735\columnwidth, trim = 0mm 0mm 0mm 0mm, clip=true]{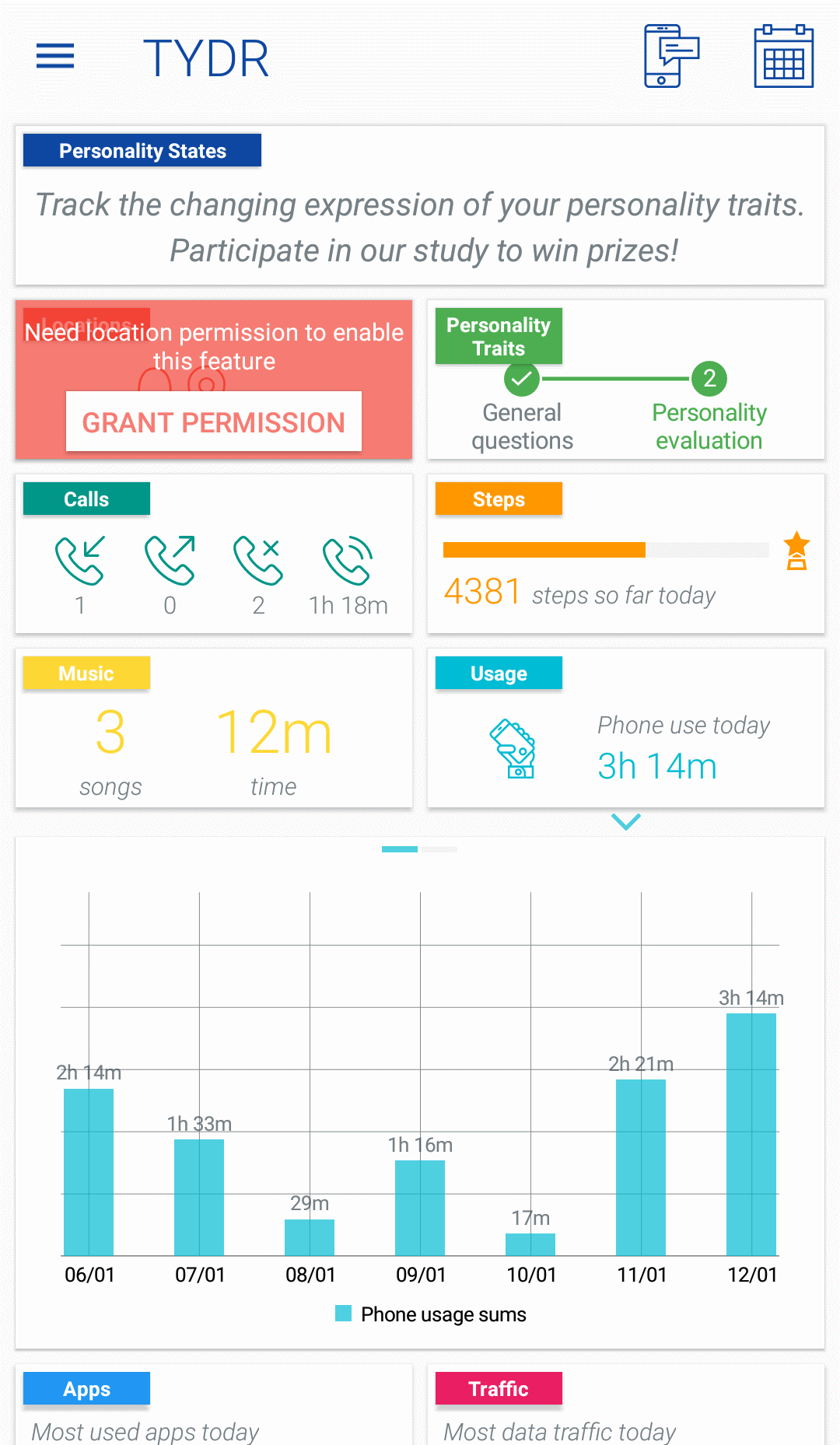}
	\caption{The main screen of TYDR that visualizes daily and weekly summaries of collected data.}
	\label{fig:main-menu}
	\includegraphics[width=0.735\columnwidth, trim = 0mm 0mm 0mm 0mm, clip=true]{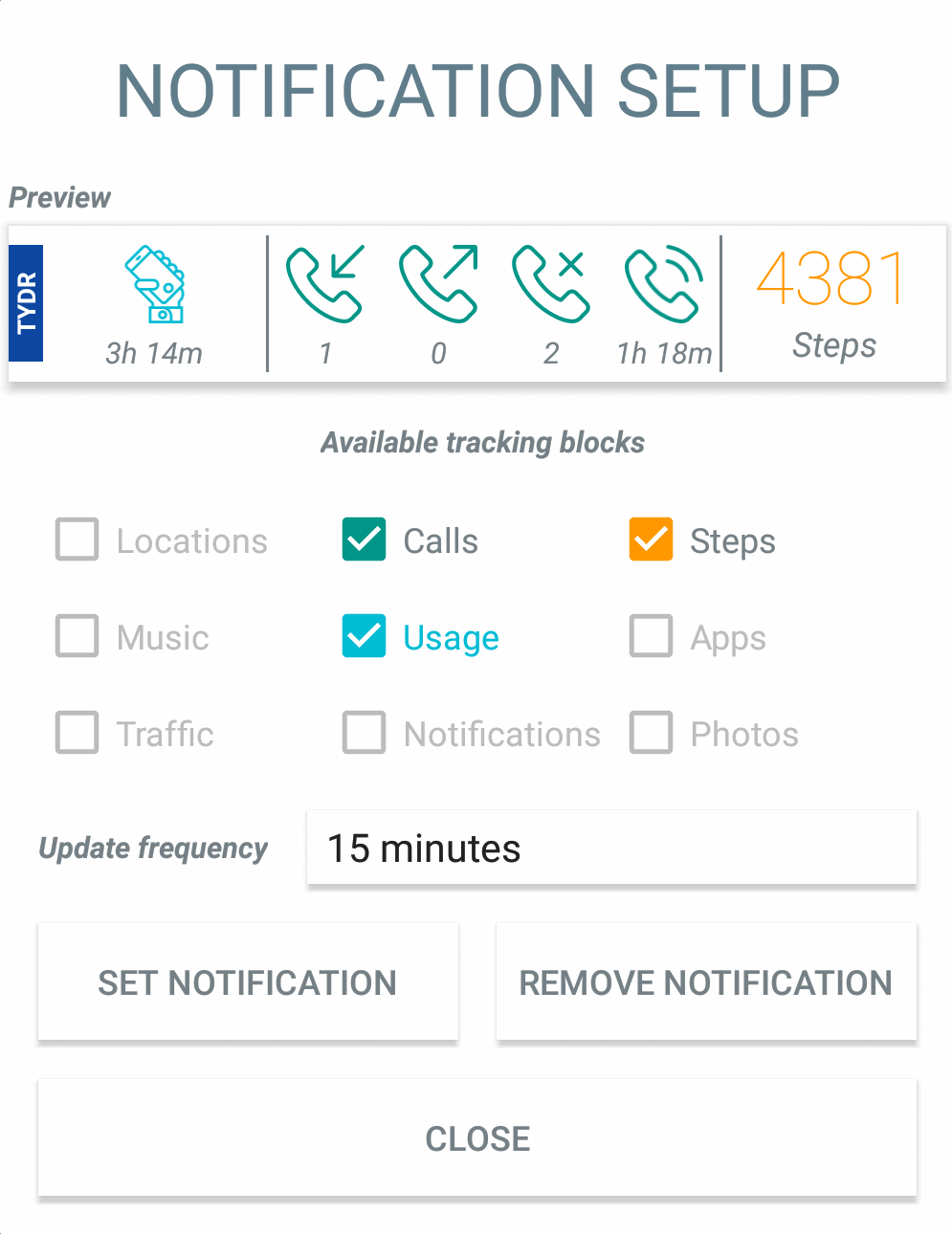}
	\caption{The user can configure the notification. The \emph{Preview} section shows what the notification will look like.}
	\label{fig:notification-configuration}
\end{figure}

The permanent notification will show up in the notification bar and
the lockscreen.
The notification is designed to be adaptive to the user's interests
by offering the possibility
to configure what information is displayed, see Figure \ref{fig:notification-configuration}.
The \emph{Preview} section in the figure shows
what the notification will look like.

\subsection{Mobile Questionnaire}
\label{sec:questionnaire}

TYDR utilizes questionnaires for demographic data that is not possible to track
automatically.
Additionally, we use standardized psychometric questionnaires 
to assess the user's personality with which we label the collected smartphone data.
To be able to update the questionnaires independently from updating
the whole app, the latest questionnaire version is fetched from the backend.

To the best of our knowledge, currently there is no official or widely adopted
library for mobile questionnaires on the Android platform.
Following general mobile survey design guidelines\footnote{E.g., \url{https://www.surveymonkey.com/mp/how-to-create-surveys/} or
\url{https://www.uxmatters.com/mt/archives/2017/02/8-best-practices-for-mobile-form-design.php}.},
we developed a questionnaire UI.
Only one question is displayed at a time, which avoids scrolling.
The users can switch between apps or turn off the screen and continue where they left off
when resuming TYDR.
A progress bar indicates how much of the current questionnaire is already filled out.
The incentive for the user to fill out the personality questionnaires
is to see their results in the related tile.

\section{Privacy Protection}
\label{sec:privacy}

In TYDR, we deal with highly sensitive data.
In this section, we detail what measures we took
in order to ensure the user's privacy.
In Section \ref{sec:priv-1}, we describe what
decisions we made during the design phase.
In Section \ref{sec:privacy-raffle},
we describe how we designed and implemented
a way to identify individual users without
linking to their data.
With this method, we are able to contact study participants that successfully
completed the study without knowing which smartphone data points are related to them.

\subsection{Privacy by Design}
\label{sec:priv-1}

The first important thing is that the app does not require any login.
This way, the user does not have to remember any login data.
Additionally, the user's data can only be stored anonymously.
However, in order to achieve our research goal to find correlations between
the user's personality and the collected smartphone data,
we need to identify which data point belongs to which user.
For this, we utilized an ID provided by Google Play Services,
that is unlikely to change during the duration of the study.
We used the identifier in salted and hashed form, in order to
further disallow any potential linking to other databases.

Furthermore, wherever possible, we store only
metadata of the sensor or usage data, for example regarding
notifications, photos, music, or calls.
Whenever we could potentially track data that would make the user
personally identifiable, we store a salted and hashed form of that
data point, for example Bluetooth device IDs or Wifi SSIDs.
Using the same salt and hash function, we still can re-identify
the same IDs without knowing the IDs themselves.
The hashing is already performed on the phone, before writing
to the local SQLite database and before uploading to our servers.

There are three main requirements the data protection official from one of our universities stated:
\begin{enumerate}
	\item Very clearly present to the user which data is being collected.
	\item Inform the users about the data, its collection, and upload before those processes of the app are started.
	\item Inform the users how they can stop the data collection and transfer.
\end{enumerate}
The first requirement is fulfilled by our privacy policy in which we list all the
data we collect and explain how it is stored and what is transferred to our servers.
The second requirement is fulfilled by the Google Play Store, where for each app,
there can be a link to the app's or developer's website and to the privacy policy.
As most users probably will not scroll down on the Google Play Store website to click and
read the privacy policy, we additionally implemented a mandatory user confirmation
of our policy in the app.
Before the app fully starts and starts collecting data, the user sees a terms and privacy policy screen.
Only after explicit confirmation, data is being collected and transferred.
The third requirement is fulfilled by informing the user
that uninstallation of the app will stop any data collection and transfer.
Additionally, we offer users the option to
contact us via a feedback form in the app to request the deletion of their data.

\subsection{Identifying Individual Users Without Linking to Their Collected Data}
\label{sec:privacy-raffle}
In psychological studies, it is common that users are compensated with university course credits,
get paid to participate, or have the chance to win money/vouchers in a raffle after study completion.
Furthermore, psychological journals -- and some computer science conferences as well -- typically require
approval codes from an ethics commission / institutional review board when submitting a paper utilizing data from a study.
Data privacy is an important factor for the approval of an ethics commission,
especially when the user data is as sensitive as the data TYDR collects.
To recognize how sensitive the collected data might be, consider, for example, that when combining
the data, we can see for a user which app he/she used for how long at which location
and what brightness the location had.
To alleviate the potential privacy concerns of users
and to receive required approvals of ethics commissions, we
developed a process that allows us to both
\begin{enumerate}
	\item check if users successfully participated in a study and
	\item contact study participants without knowing which smartphone data belongs to them.
\end{enumerate}
This way, we can directly draw and contact the winners of a raffle,
without having a link between collected data and meaningful user identifiers, i.e.,
email addresses.
Additionally generated participation codes can further be used for
claiming university credit points, if applicable.
There are three steps to our design of this system:

\textbf{User sign-up.}
In order to participate in the study, additionally to just installing the app,
users have to sign up with an email address, so we can contact the winners of the raffle.
Entering an email address has the potential to de-anonymize the data.
We store the data about the study participants in a separate table in the backend
that is not linked to the tables containing smartphone data.

\textbf{Checking study participation success.}
The planned study includes the commitment of the participants
to fill out daily psychological questionnaires.
Users who do not fill those out regularly should not be eligible to receive
compensation.
As we have no link between the identifiable study participants and their smartphone
data, we cannot check the rate of filling out the daily questionnaire on the backend.
However, we can check this rate in the app and report to the backend whether it meets the required percentage.
	
\textbf{Generating participation codes.}
After successful study completion, the user can trigger the study completion by
pressing the related button in the sidebar menu.
This triggers the app to let the backend know about the successful completion of the study.
The backend then generates an individual participation code and
replies to the app's query with it.
This code can then be used to
indicate
successful participation to, e.g.,
claim university credit points.

\section{Conclusion and Future Work}
\label{conclusion}

In this paper, we presented the design and implementation
of TYDR, an Android app which tracks the user's
smartphone sensor and usage data and queries the user
with personality questionnaires.
We will utilize this app for conducting a study
with the goal to predict the user's personality from smartphone data.
We detailed the challenges in implementing such a system.
Sensor data has to be tracked efficiently by finding the right
balance between the amount of data to be stored, battery consumption,
and level of detail of the information.
The user interface should be attractive and intuitive.
We turned the data collection into TYDR's core feature of
processing and visualizing daily statistics for the user.
Especially when dealing with sensitive data like in TYDR,
privacy and data anonymization have to be a key concern.
We showed how we designed our app with privacy in mind,
taking into account the concrete requirements that are posed
by data protection officials and ethics commissions / institutional review boards.
Furthermore, we showed how we
can identify individual users that successfully completed a study without
linking their email addresses to the smartphone data we collected from them.
The process for this consists of storing contact data separately from
the collected data and letting the app check the requirements for successful
study completion.

Future work includes conducting the planned study.
Both the described implementation techniques for
smartphone sensor and usage data tracking as well
as the results from the future study can be incorporated
in a variety of applications.
This includes, for example, applications related to mobile health
\cite{PryssMobileCrowdSensing2015} or
mobile social networking \cite{beierle_towards_2015}.

\begin{acks}
This work was done in the context of
project DYNAMIC\footnote{\url{http://www.dynamic-project.de}} (grant No 01IS12056),
which is funded
as part of the Software Campus initiative by the German
Federal Ministry of Education and Research (BMBF).
We are grateful for the support provided by
Soumya Siladitya Mishra,
Sarjo Das,
Oksana Kalynchuk,
Michael Segert,
Parth Singh, and
Bernd Louis.
\end{acks}

\bibliographystyle{ACM-Reference-Format}

\end{document}